\documentclass{eas}
\usepackage{graphicx}
\newcommand{\rp}{r_{\mathrm{p}}}

\newcommand{\lsim}{\lower0.6ex\vbox{\hbox{$ \buildrel{\textstyle <}\over{\sim}\ $}}}
\newcommand{\gsim}{\lower0.6ex\vbox{\hbox{$ \buildrel{\textstyle >}\over{\sim}\ $}}}

\newcommand{\acost}{\vert \cos(\theta)\vert}
\newcommand{\Pks}{P_{\mathrm{KS}}}

\newcommand{\Vsat}{V_{\mathrm{max}}^{\mathrm{sat}}}
\newcommand{\Vhost}{V_{\mathrm{max}}^{\mathrm{host}}}

\newcommand{\kms}{\, \mathrm{kms}^{-1}}
\newcommand{\hMsun}{\, h^{-1}\mathrm{M}_{\odot}}
\newcommand{\hpc}{\, h^{-1}\mathrm{pc}}

\newcommand{\fsub}{f_{\mathrm{sub}}}

\begin{document}

\TitreGlobal{Mass Profiles and Shapes of Cosmological Structures}

\title{
The Anisotropic Spatial Distribution of CDM Subhalos
}
\author{
Andrew R. Zentner
}
\address{
Kavli Institute for Cosmological Physics and 
Department of Astronomy and Astrophysics, 
The University of Chicago, Chicago, IL 60637 USA
}
%
\runningtitle{Anisotropy of CDM Subhalo Distributions}
\setcounter{page}{23}
%
\index{
Andrew R. Zentner
}
%
%
%
%
%
\begin{abstract} 
I review recent results on the relative spatial distribution of 
substructure in CDM halos.  I show that the spatial distribution 
of subhalos is anisotropic and generally prolate with a long axis 
that is closely aligned with 
the long axis of the mass distribution of 
the host halo.  I show that the correlation between the 
subhalo distribution and the long axis of the host halo 
is strong both in dissipationless and 
dissipational gasdynamical simulations.  
More massive subhalos tend to be more strongly 
clustered along the major axis of the host halo reflecting 
filamentary accretion.  The anisotropy of subhalos 
has potential implications for the interpretation of several 
observations both in the Local Group and beyond.  
For example, I show that while the mean 
projected mass fraction in substructure in the central 
regions of CDM halos is $\fsub \approx 0.4\%$, 
$\fsub$ is a strong function of projection angle and 
is $\sim 6$ times higher for projections 
nearly collinear with the major axis of the host.  
\end{abstract}
\maketitle
%
%

\section{The Planarity of Milky Way Satellites}

Holmberg (1969) reported that satellite galaxies of spiral 
primaries with projected separations $\rp \lsim 50$~kpc 
are tend to lie near the short axes of the light 
distributions of their primaries.  
Zaritsky et al. (1997) revisited this issue and 
found evidence for alignment in the same 
sense as Holmberg for satellites within 
$\rp \sim 200-500$~kpc.  The 
angular distribution of galaxies has been 
of recent interest with increased data from 
large surveys and the possibility of these data 
to relate the orientations of galaxies to their 
halos in a statistical way 
(e.g., Sales \& Lambas 2004; Brainerd 2004; 
Azzaro et al. 2005, A05), as well as 
studies of satellites in the Local Group.

Kroupa et al. (2005, K05) recently argued that 
the nearly planar distribution of Milky Way 
(MW) satellites is a serious challenge to 
the standard cold dark matter (CDM) paradigm 
of structure formation.  
Zentner et al. (2005, Z05) addressed this 
issue from a theoretical standpoint, and 
similar results were reported in contemporaneous 
papers Libeskind et al. (2005, L05) and Kang et al. (2005).  
They showed that the conclusions of K05 
were incorrect for two reasons:  first, the 
statistical analysis of K05 was not valid for 
small samples, such as the $11$ observed MW 
satellites, and for such samples the statistic 
they used is non-discriminatory; second, K05 
incorrectly assumed that the null hypothesis 
for CDM should be an isotropic satellite 
distribution.

%
%
\begin{figure}[t]
\centering
\includegraphics[width=11cm]{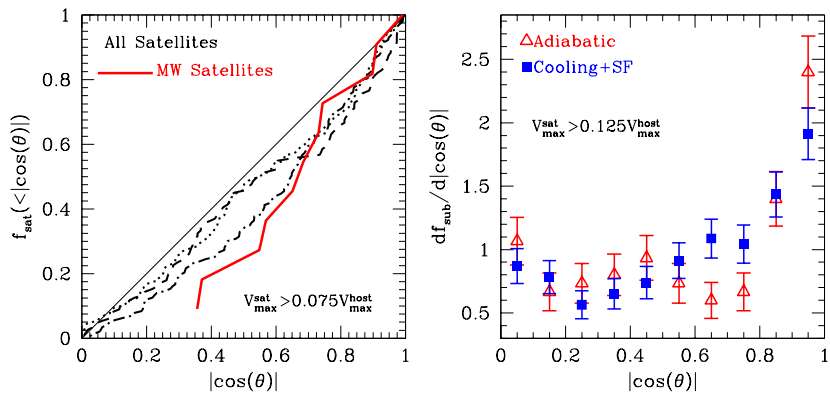}
\caption{
{\it Left:}  
The cumulative fraction of satellites with an 
angular position $<\acost$ from the major axis of the 
host halo mass distribution as a function of $\acost$.  
The {\em thin, solid} line represents an isotropic 
distribution.  The {\em dashed, dotted}, and 
{\em dot-dashed} lines are the distributions of 
subhalos of $3$ simulated MW host halos, 
with $\Vsat \ge 0.075\Vhost$.  
The {\em thick, solid} line represents the $11$  
MW satellites.  The MW satellites 
are placed on the plot by {\em assuming} 
that the rotation axis of the MW is aligned with 
the major axis of the halo.
{\it Right:}  
The differential fraction of subhalos as a 
function of angular displacement from the major axis 
of the primary, cluster halo.  An isotropic 
distribution is uniform in $\acost$.  
The {\em triangles} show the results from $8$ dissipationless 
cluster simulations employing adiabatic gas physics.  
The {\em squares} show results from simulations 
of the same $8$ clusters including radiative gas 
cooling and star formation.
}
\label{f1}
\end{figure}
%

Z05 showed that the distribution of CDM subhalos 
is anisotropic.  Subhalos or subsets thereof, 
the likely sites of galaxy formation, 
are preferentially aligned near the long 
axes of the triaxial mass distributions of 
their primary halos.  
This is shown explicitly in the left panel of 
Figure~\ref{f1} for a sample of $3$ simulated 
approximately MW-sized, 
CDM halos (see Z05 for details).  
The angle between the major axis of the primary halo 
and the position of the 
subhalo is $\theta$ and an isotropic 
distribution is uniform in the variable $\cos(\theta)$.  
The principal axes of the host halo were 
computed using only particles 
within $30\%$ of the halo virial radius, 
to focus on the region where the 
central galaxy resides.  The satellites were selected 
to have maximum circular velocities $\Vsat \ge 0.075\Vhost$, 
where $\Vhost$ is that of the host halo.  
These satellites are roughly the size required 
to host the observed MW dwarf satellites 
(e.g., Kravtsov et al. 2004).  
The Kolmogorov-Smirnov probability of 
selecting the simulated subhalo sample 
from an isotropic distribution is 
$\Pks \sim 10^{-4}$.

In addition, Z05 demonstrated that planar 
distributions of subhalos, 
similar to that of the 
MW satellites, are not unlikely due largely 
to accretion along preferred directions. Such 
planes are typically aligned with the major 
axis of the primary halo.  Thus, the MW satellites 
are consistent with CDM predictions, provided that 
the pole of the MW is aligned with the major 
axis of the surrounding halo.  The metal-poor 
globular clusters surrounding the MW and the 
satellites of M31 show evidence of a similar 
alignment and new techniques may yield constraints 
on the orientation of the MW halo 
(e.g., Gnedin et al. 2005).  
However, such alignments present a 
challenge for simple scenarios of disk galaxy 
formation because the angular momenta of DM halos 
tend to be perpendicular to halo major axes.

The results of Z05, Kang et al. (2005) and L05 
are all based on dissipationless 
$N$-body simulations; 
however, one of the effects of baryonic 
dissipation is to make DM halos more spherical than 
their counterparts in dissipationless simulations 
(e.g., Kazantzidis et al. 2004).  One may inquire 
whether the alignment of satellites along the principal 
axes of host halos is as prevalent in 
dissipational simulations.  
One might expect differences between 
dissipational and dissipationless simulations 
to be small in this regard because 
both the major axis and the positions of 
satellites reflect the directions of 
recent accretion along filaments and because subhalos 
are biased toward large halo-centric 
distances compared to DM ($r \gsim 0.3R_{\mathrm{vir}}$), 
where the change in shape is small.  
The right panel of Figure~\ref{f1} 
is an explicit demonstration that 
dissipational processes do not 
significantly alter the alignment of halo 
and satellites.  The figure shows an analysis 
of the eight cluster halos 
of Kazantzidis et al. (2004) simulated once with  
dissipationless, adiabatic gas physics and a 
second time including radiative cooling and 
star formation.  Though the inner halos in the 
cooling simulations are significantly rounder, the 
alignment of host halo and satellites remains 
pronounced.

%
%
%
%
%

\section{Is There an Angular Bias Between Subhalos and Dark Matter?}

%
%
%
%
\begin{figure}[t]
\centering
\includegraphics[width=11cm]{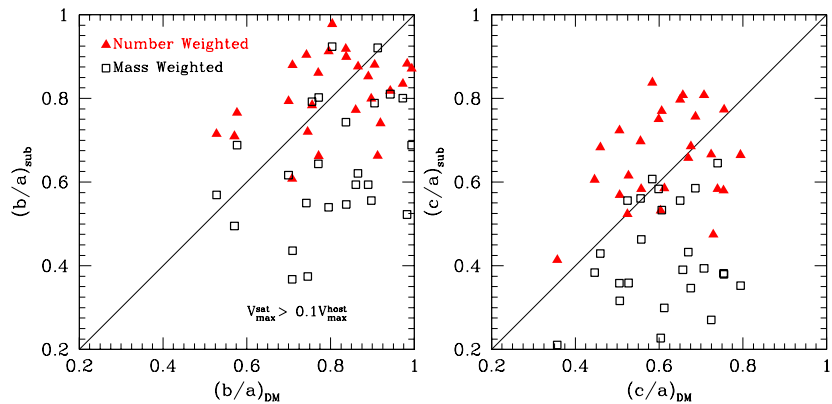}
\caption{
The shape distribution of DM compared to the 
distribution of satellite halos.  The {\em left} 
panel shows the axis ratio $(b/a)$, while the 
{\em right} panel shows the axis ratio $(c/a)$.  
Each panel is a scatter plot of the axis ratios 
of {\em all} DM in each host halo on the 
{\em horizontal} axis against the axis ratios 
of each of the subhalo populations on the 
{\em vertical} axis.  The {\em triangles} show 
number-weighted subhalo axis ratios and the 
{\em squares} represent mass-weighted subhalo 
axis ratios.  All subhalos with 
$\Vsat \ge 0.1\Vhost$ are included.  
}
\label{f2}
\end{figure}
%
%

It is interesting to quantify the relationship between 
the spatial distributions of the smooth, 
DM components of host halos and the 
subhalos that reside within them.  Do the subhalos simply 
follow the triaxial mass distribution?  There are 
several potential ways to address this issue, such 
as computing angular correlations etc., and I discuss 
two intriguing quantifications in this section.  
One way to address the relationship of 
subhalos and DM is through the ratios 
of the principal axes of inertia denoted $a \ge b \ge c$.  
For subhalos, the inertia tensor can be computed in two 
ways.  In the first, each subhalo is counted 
equally and in the second method, each subhalo 
can be counted in proportion to its bound mass.  
The result is a ``number-weighted'' inertia tensor 
and a ``mass-weighted'' inertia tensor.  Figure~\ref{f2} 
shows a comparison between the axis ratios of host DM 
halos using, computed as specified above, 
and the mass- and number-weighted axis ratios of their  
subhalo populations.  The sample consists of 
$26$ hosts with $180\kms \le \Vhost \le 400 \kms$ 
and their subhalos simulated with the ART code 
(Kravtsov et al. 1997).  The particle mass is 
$m_{\mathrm{p}} = 4.9 \times 10^6 \hMsun$, 
the spatial resolution is $\sim 150 \hpc$, 
and each host contains $\gsim 2\times 10^5$ 
particles within its virial radius.

The number-weighted axis ratios in 
Fig.~\ref{f2} show that the full number-weighted 
satellite populations broadly trace the DM 
distributions of their host halos.  However, notice that the 
mass-weighted axis ratios are systematically 
smaller than that of the DM in the host halo.  
More massive subhalos are more strongly biased toward 
a flattened distribution than small subhalos, a 
result consistent with the studies of 
Z05, L05, and A05.  The robustness of this result has 
been checked by randomly re-assigning the weights 
(masses in this case) among the subhalo populations.  
The axis ratios based on these randomized weights are 
generally similar to the number-weighted axis ratios 
shown in Figure~\ref{f2}.  This angular bias for large 
subhalos is not entirely surprising.  
The smallest subhalos have generally 
been accreted over an extended period of time 
and interact gravitationally as DM particles, 
adopting a self-consistent configuration with the host 
potential.  The largest subhalos have typically been 
accreted more recently so they more faithfully reflect 
the directions of recent infall, and they tend to be 
more strongly biased to form in overdense filaments.

\begin{figure}[t]
\centering
\includegraphics[width=6cm]{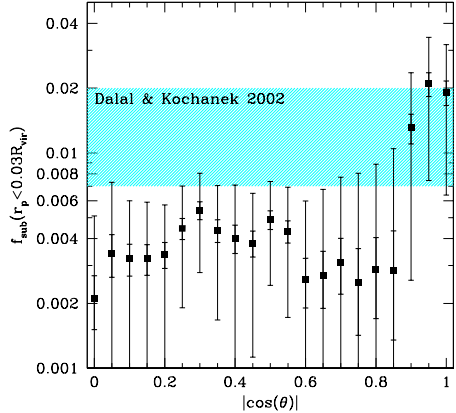}
\caption{
Substructure mass fractions as a function of 
projection angle from the major axis of the host 
halo.  The {\em squares} represent the average 
$\fsub$ measured from all $26$ host halos, using 
two projections for each host.  
The {\em outer} errorbars represent the 
scatter among projections and the 
{\em inner} errorbars represent the estimated 
error in the mean of $\fsub$.  The shaded band 
represents the $90$\% confidence region of 
$\fsub$ from the measurement of 
Dalal \& Kochanek (2002).  
}
\label{f3}
\end{figure}

As a second comparison between substructure and 
smooth mass, consider the 2D projected fraction of 
mass in substructure, $\fsub$.  This quantity is 
constrained by measurements of flux ratio anomalies 
in multiply-imaged quasar systems (e.g., Dalal \& Kochanek 2002), 
and can be used as a probe of cosmological parameters 
that influence the growth of small-scale structure.  
Zentner \& Bullock (2003) and Mao et al. (2004) have 
made predictions for the mean projected substructure mass 
fractions, with the former considering a variety of 
primordial power spectra and dark matter properties.   
In what follows I show the projected substructure 
mass fraction as a function of 
projection angle $\theta$, from the major axis of the 
host halo.  Following Mao et al. (2004), I have 
computed $\fsub(\theta)$ as a function of projection angle 
using all mass and substructures within $3$ 
virial radii of the center of each host, 
in order to include correlated material associated 
with each halo.  I projected in cylinders of 
radius $\rp = 0.03R_{\mathrm{vir}}$, comparable 
to the Einstein radii of strong-lens 
systems.  The result is shown in Figure~\ref{f3}, 
along with the observed $90$\% 
confidence region for $\fsub$ measured in 
quadruply-imaged systems by Dalal \& Kochanek (2002).  
The mean substructure mass fraction is approximately 
$\fsub \approx 0.4$\% with a large scatter, 
consistent with Mao et al. (2004).  
Interestingly, $\fsub(\acost)$ is 
$\sim 5-6$ times higher for projections near the 
long axis of the host.  If elliptical galaxies 
are well aligned with their host halos, this 
result may have important consequences for 
determinations of $\fsub$ in multiply-imaged 
quasar systems and several other observed 
properties of strong lenses.

%
%
%
%

%
%
%
%


\newpage

\acknowledgements

I am grateful to my collaborators Brandon Allgood, 
Oleg Gnedin, Anatoly Klypin, Andrey Kravtsov, Daisuke Nagai, 
and Eduardo Rozo for their invaluable contributions to this 
research.  I thank James Bullock, Neal Dalal, 
Stelios Kazantzidis, Chuck Keeton, Ben Metcalf, 
and Jeremy Tinker for helpful discussions.  
ARZ is funded by The Kavli Institute 
for Cosmological Physics at 
The University of Chicago and by the 
National Science Foundation 
under grant No. NSF PHY 0114422.

\endacknowledgements

%
%
%


\begin{thebibliography}{}


%
%
%


\bibitem[Azzarro et al.(2005)]{azzaro_etal05}
Azzaro, M., Zentner, A.~R., Prada, F., \& Klypin, A.~A., 2005, 
ApJ Submitted (astro-ph/0506547)

\bibitem[Brainerd(2004)]{brainerd04}
Brainerd, T.~G., 2004, ApJL Submitted (astro-ph/0408559)

\bibitem[Dalal \& Kochanek(2002)]{dalal_kochanek02}
Dalal, N. \& Kochanek, C.~S., 2002, ApJ, 572, 25


\bibitem[Gnedin et al.(2005)]{gnedin_etal05}
Gnedin, O.~Y. et al., 2005, ApJ In Press (astro-ph/0506739)


\bibitem[Holmberg(1969)]{holmberg69}
Holmberg, E., 1969, Arkiv Astron., 5, 305

\bibitem[Kang et al.(2005)]{kang_etal05}
Kang, X., Mao, S., Gao, L., \& Jing, Y.~P., 2005, 
A\&A Accepted

\bibitem[Kazantzidis et al.(2004)]{kazantzidis_etal04} 
Kazantzidis, S. et al., 2004, ApJL, 611, L73

%
%

\bibitem[Kravtsov et al.(2004)]{kravtsov_etal04}
Kravtsov, A.~V., Gnedin, O.~Y., and Klypin, A.~A. 2004, 
ApJ, 609, 482

\bibitem[Kravtsov et al.(1997)]{kravtsov_etal97}
Kravtsov, A.~V., Klypin, A.~A., \& Khokhlov, A.~M., 1997, 
ApJS, 111, 73

\bibitem[Kroupa, Theis, \& Boily(2005)]{kroupa_etal05}
Kroupa, P., Theis, C., \& Boily, C.~M., 2005, A\&A, 431, 517

\bibitem[Libeskind et al.(2005)]{libeskind_etal05} 
Libeskind, N.~I. et al., 2005, 
MNRAS, Submitted (astro-ph/0503400)

\bibitem[Mao et al.(2004)]{mao_etal04}
Mao, S., Jing, Y.~P., Ostriker, J.~P., \& Weller, J., 
2004, ApJL, 604, L5

\bibitem[Sales \& Lambas(2005)]{sales_lambas05}
Sales, L. \& Lambas, D.~G., 2005, MNRAS, 310, 1087

\bibitem[Zaritsky et al.(1997)]{zaritsky_etal97}
Zaritsky, D., Smith, R., Frenk, C.~S., \& White, S.~D.~M., 1997, 
ApJ, 478, 39

\bibitem[Zentner \& Bullock(2003)]{zentner_bullock03}
Zentner, A.~R., \& Bullock, J.~S., 2003, 
ApJ, 598, 49

\bibitem[Zentner et al.(2005)]{zentner_etal05b} 
Zentner, A.~R., Kravtsov, A.~V., Gnedin, O.~Y., 
\& Klypin, A.~A., 2005, ApJ, 629, 219

\end{thebibliography}
\end{document}